\documentclass[12pt]{article}
\usepackage{bm}
 \usepackage{graphicx}
 
\begin{document}

\begin{center}
{\large\bf On Alfv\'en hypothesis about nuclear hydromagnetic resonances}

\vspace{1.cm}

{S. I. Bastrukov$^{a,b}$, I. V.  Molodtsova$^{c}$,  J. W. Yu$^{a}$, R. X. Xu$^{a}$}

\vspace{0.5cm}

{\it $^a$State Key Laboratory of Nuclear Physics and Technology,\\
 School of Physics, Peking University, 100871 Beijing, China\\
 $^b$ Data Storage Institute, A*Star, Singapore 117608, Singapore\\
 $^c$ Joint Institute for Nuclear Research, Dubna 141980, Russia}

\vspace{.5cm}

\begin{abstract}
 The atomic nucleus capability of responding by hydromagnetic vibrations, 
 that has been considered long ago by  Hannes Alfv\'en, is re-examined in 
 the context of current development of nuclear physics and pulsar astrophysics. 
 \\[0.2cm]
\end{abstract}

\end{center}

\newpage 

\section{Introduction}
  In the long-ago published article\cite{A-57}, Hannes Alfv\'en pointed out that electromagnetic 
  response of an atomic nucleus should manifest features generic in 
  hydromagnetic vibrations of an ultra fine piece of a perfectly conducting continuous medium of nuclear density 
  with frozen-in magnetic field and made an attempt to evaluate "the order of magnitude of possible magnetohydrodynamic (MHD) resonance 
  frequencies"\footnote{By the 
  time of publication of  Alfv\'en work\cite{A-57}, all attempts to 
  understand macroscopic properties of nuclear matter, regarded as a continuous medium, have 
  been dominated by Gamow's idea about similarity 
  between an atomic nucleus and a drop of liquid mercury that has been laid  
  at the base of the nuclear liquid drop model. This similarity has been 
  used as a guide in obtaining semi-empirical formula for nuclear binding 
  energy\cite{RS-94} and, most extensively, in first macroscopic, electro-capillary, theory of 
  nuclear fission\cite{FR-74}. From the history of magnetohydrodynamical 
  investigations\cite{Ch-61,AF-63,F-07} it is known that the liquid mercury is 
  the material in which hydromagnetic waves have first been discovered in widely 
  known Lundquist's experiments. With this in mind, it seems quite plausible that all the 
  above have led Alfv\'en to suggest that atomic nucleus can too respond by 
  hydromagnetic vibrations whose exciation presumes the presence in the nucleus of frozen-in magnetic field.
  Together with this, it seems worth noting that while 
 the MHD investigations have a long story, the
 consistent theory of  vibrational MHD modes in a spherical mass has only recently been substantially developed. The extensive discussion of this theory in the context of asteroseismology of neutron stars can be found in [8-15] 
 and in context of physics of nano-particles in [16-18].}.
  In this communication we revisit this Alfv\'en proposition with focus 
  on the magnitude of intranuclear magnetic field, whose presence in the nucleus volume 
  is the chief prerequisite of sustaining MHD oscillations. In approaching this issue it seems best 
  to start with the current understanding of "the history of matter from big 
  bang to the present"\cite{AR-96}, which teaches us that the nuclear material objects, both neutron stars and atomic nuclei heavier 
  than Fe-56, are produced in the magnetic-flux-conserving core-collapse supernovae.
  The implosive contraction of massive main-sequence star gives birth to a neutron star - neutron-dominated mass of 
  self-gravitating nuclear matter with frozen-in magnetic field of extremely large intensity. 
  Since the r-process of explosive nucleosynthesis proceeds too in the presence of super strong magnetic field, it seems not inconsistent 
  to expect that neutron-dominated mid-weigth and heavy nuclei come, like neutron stars, into existence with entrapped magnetic field.
  In other words, the frozen-in magnetic field is the fundamental property of both neutron stars and heavy atomic nuclei.

  Before embarking on theoretical underpinning for the collective model of 
 nuclear hydromagnetic vibrational response, we remind that 
 basic purpose of continuum-mechanical description of nuclear giant resonances in terms of vibrational eigenstates of an 
 ultra-fine piece of continuous nuclear matter is to gain some insight into macroscopic properties of nuclear material.
 The macroscopic nature of giant resonance is determined  by restoring force. The position of energy centroid of a resonance 
   in the nuclear spectrum is defined by the standard quantum-mechanical equation 
   $E_{GR}=\hbar\omega$, where $\omega$ is the frequency of nuclear 
   vibrations carrying information about electromagnetic and mechanical parameters of 
   nuclear material and upon the nucleus radius $R=r_0A^{1/3}$. The key idea is 
   to extract the magnitude of these parameters by identifying 
   theoretical and empirical energies of resonance under consideration. 
   A representative example of such an approach is the 
   macroscopic treatment of isoscalar giant resonances  in terms 
   of spheroidal and torsional modes of shear elastic vibrations of a solid sphere
   whose fundamental frequency reads $\omega_e=[\mu/(\rho 
    R^2)]^{1/2}$,  where $\mu$ is the shear modulus of nuclear matter.
   This interpretation rests on observation 
   \cite{B-74} that the energy of vibrational eigenstates,
   $E_e\sim \hbar\omega_e\sim A^{-1/3}$, has one and the same 
   mass-number dependence as the empirical energy of giant isoscalar resonances 
   $E_{GR}\sim  A^{-1/3}$.  The main outcome of this line of argument 
   consists in an assessment of shear modulus $\mu$ of nuclear 
   material,  $10^{33} <\mu<10^{34}$ dyn cm$^{-2}$, which is of 
   particular interest in the asteroseismology of neutron stars (e.g., [22-24]
   and references therein). In this context it worth mentioning that magneto-hydrodynamic theory rests on the statement 
   that magnetic field pervading perfectly conducting medium imparts to it a supplementary portion of 
   solid-mechanical elasticity\cite{Ch-61,AF-63}. This suggests that hydromagnetic vibrations in question
   should have some features in common with elastic vibrations of solid sphere and, hence, manifest itself 
   as giant resonances of isoscalar type.
   Adhearing to the idea that the mid-weight and heavy nuclei 
   are produced (in r-process of explosive nucleosynthesis) with frozen-in magnetic field, we consider the nuclear 
   MHD vibrations with focus not on the energy of hydromagnetic 
   resonances but on the intensity of intranuclear magnetic field. Namely, having observed 
   that the mass-number dependence of energy of hydromagnetic resonant excitations is similat to 
   that for empirically established giant resonances 
   we show that above line of reasoning allows one to evaluate the magnetic field magnitude.

\section{Governing equations}  
  
Following a line of Alfv\'en's argument\cite{A-57}, we assume that strongly collective 
response of atomic nucleus (to perturbation induced by inelastically scattered electrons or elastically scatted photons) is dominated by hydromagnetic vibrations. 
 The relevant to this case MHD equations can be conveniently written in the form\cite{Ch-61}
   \begin{eqnarray}
  \label{e1.1}
 && \rho\delta{\dot {\bf v}}=\frac{1}{c}
 [\delta {\bf j}\times {\bf B}],\quad \delta {\bf j}=\frac{c}{4\pi}[\nabla\times \delta {\bf B}],\\
  \label{e1.2}
 && \delta {\dot {\bf B}}=\nabla\times [\delta {\bf v}\times {\bf B}], \quad \nabla\cdot \delta {\bf v}=0
  \end{eqnarray}
  These equations describe Lorenz-force-driven oscillations 
  of velocity $\delta {\bf v}$ of material flow coupled with fluctuations of  
  magnetic field $\delta {\bf B}$ about immobile equilibrium state of 
  incompressible and perfectly conducing continuous medium of density $\rho$ 
  pervaded by magnetic field ${\bf B}$.  
  Taking into account that $\delta {\bf v}={\dot
  {\bf u}}$ where ${\bf u}$ is the field of material displacement (which is
  the basic variable of solid-mechanical theory of elasticity), the coupled 
  equations (\ref{e1.1}) and (\ref{e1.2}) can be reduced to only 
  one equation\footnote{
  The above mentioned analogy between oscillatory behavior of perfectly conducting medium pervaded 
  by magnetic field (magneto-active plasma) 
  and elastic solid, regarded as a material continuum, is 
  strengthened by the following tensor representation of the last equation
   $\rho\,{\ddot u}_{i}=\nabla_k \delta M_{ik}$, where $\delta M_{ik}=(1/4\pi)
   [B_i\delta 
  B_k+B_k\delta B_i-B_j\delta B_j\delta_{ik}]$ is the Maxwellian tensor of  
 magnetic field stresses with  $\delta B_i = \nabla_k[u_i B_k - u_k B_i]$. 
 This form is identical in appearance to canonical equation of solid-mechanics 
 $\rho\,{\ddot u}_{i}=\nabla_k \sigma_{ik}$, where
 $\sigma_{ik}=2\mu\,u_{ik}+[\kappa-(2/3)\mu]\,u_{jj}\delta_{ik}$ is the
 Hookean tensor of mechanical stresses and $u_{ik}=(1/2) [\nabla_{i}\,u_k+\nabla_k\,u_i]$ is the 
  tensor of shear deformations in an isotropic elastic continuous matter with shear
  modulus $\mu$ and bulk modulus $\kappa$ (having physical dimension of 
  pressure). 
  }
     \begin{eqnarray}
  \label{e1.3}
  && \rho\, {\ddot {\bf u}}=\frac{1}{4\pi}
 [\nabla\times[\nabla\times [{\bf u}\times {\bf B}]]\times {\bf B}].
  \end{eqnarray} 
 In approximation of node-free vibrations, widely used in macroscopic models of 
    collective nuclear 
    dynamics,  the frequency of Alfv\'en hydromagnetic modes can be computed 
    by the energy method which rests on integral equation  
    of energy balance  
    \begin{eqnarray} 
\label{e1.4}
\frac{\partial }{\partial t}\int\frac{\rho {\dot {\bf u}}^2}{2}d{\cal V}=\frac{-1} 
{4\pi}\int [{\bf B}\times
 [\nabla\times[\nabla\times [{\bf u}\times {\bf B}]]]\cdot {\dot {\bf u}}d{\cal V}
   \end{eqnarray}        
   In this method, the bulk density $\rho$ and the shape of frozen-in 
   magnetic field ${\bf B}$ are regarded as known functions of position, 
   $\rho(r)=\rho\,f(r)$ and ${\bf B}({\bf r})=B\,{\bf b}({\bf 
   r})$, where $\rho={\rm constant}$ is the density in the nucleus center and 
   dimensionless scalar function $f(r)$ describes the density profile, by $B={\rm 
   constant}$ is denoted the magnetic field intensity and ${\bf b}({\bf r})$ stands for 
   the dimensionless vector-function of spatial distribution of the field over the nucleus 
   volume.  Substitution in the latter equation of the following separable 
   representation of fluctuating material displacements 
  \begin{eqnarray}
\label{e1.5} 
 &&{\bf u}({\bf r},t)={\bf a}({\bf r})\,\alpha(t),
 \end{eqnarray}
 where ${\bf a}({\bf r})$ is the time-independent field of instantaneous 
 displacements, leads to equation for amplitude $\alpha(t)$ describing 
 harmonic vibrations  
 \begin{eqnarray}
 \label{e1.6}
  &&\frac{d{\cal H}_A}{dt}=0,\quad {\cal H}_A=\frac{{\cal M}{\dot \alpha}^2(t)}
  {2}+
  \frac{{\cal K}\alpha^2(t)}{2},\\
  \label{e1.7}
  &&{\cal M}{\ddot \alpha}(t)+{\cal K}\alpha(t)=0,\quad {\cal M}=\rho\,m, \quad  
  {\cal K}=\frac{B^2}{4\pi}\, k,\\
   \label{e1.8}
  && m=\int f(r)\,{\bf a}({\bf r})\cdot {\bf a}({\bf r})\,d{\cal V},\\
   \label{e1.9}
  &&  k=\int
 {\bf a}({\bf r})\cdot[{\bf b}({\bf r})\times[\nabla\times[\nabla\times[{\bf a}({\bf
 r})\times{\bf b}({\bf r})]]]]\,d{\cal V}.
 \end{eqnarray}
 The general analytic expression for the spectrum of discrete frequencies of 
 MHD oscillations, $\omega_\ell(MHD)$, can be represented as 
 follows 
 \begin{eqnarray}
  \label{e1.10}
  && \omega_\ell(MHD)=\sqrt{\frac{\cal K}{\cal M}}=\omega_A(B)\,s_\ell,\quad \omega_A(B)=\frac{v_A}{R},\quad v_A=\frac{B}{\sqrt{4\pi\rho}}
  \end{eqnarray}
  where $\omega_A(B)$ stands for the Alfv\'en frequency which is the natural unit of frequency of MHD oscillations depending 
  only on the field strength $B$ 
  and $s_\ell$ is numerical spectral factor depending of multipole degree $\ell$ of hydromagnetic oscillations.
  As illustrative example relevant to the subject of this work, we present the result of 
  calculations with frozen-in magnetic field of an axisymmetric configuration, pictured in Fig.1, whose spherical components are  
  \begin{eqnarray}
  \label{e1.11}
  {\bf b}=\left[b_r=0,\, b_\theta=0,\,b_\phi(r,\theta)=\frac{(R^2-r^2\sin^2\theta)^{1/2}}{R}\right]. 
  \end{eqnarray} 
   The frequency spectrum of MHD vibrations with the node-free irrotational field of instaneneous displacements 
   (as is the case of nuclear giant resonances of electric type) 
  \begin{eqnarray}
  \label{e1.12} 
  {\bf a}=A_\ell\nabla [r^\ell P_\ell(\theta)] 
  \end{eqnarray}
  (where $P_\ell(\theta)$ being Legendre polynomial of multipole degree $\ell$) is 
  given by 
  \begin{eqnarray}
  \label{e1.13}
  \omega_\ell(MHD)=\omega_A(B)\,s_\ell,\quad s_\ell=\left[\frac{(2\ell+1)(\ell-1)}{(2\ell-1)}\right]^{1/2}.
  \end{eqnarray}
  
   \begin{figure}
 \centering{\includegraphics[width=8.0cm]{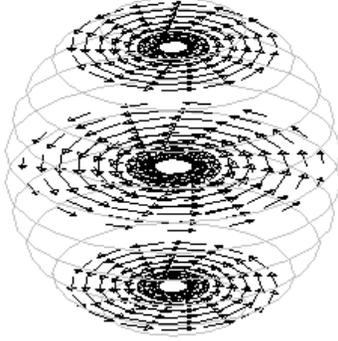}}
 \caption{An illustrative example of lines of frozen-in magnetic field having axisymmetric pure toroidal 
 configuration: ${\bf B}_{t}=[B_r=0,\, B_\theta=0,\,b_\phi(r,\theta)=(B/R)[R^2-r^2\sin^2\theta]^{1/2}$ with $B={\rm constant}$.}
\end{figure}

  The basic frequency of Alfv\'en oscillations $\omega_A$ of a spherical mass $M=(4\pi/3)\rho R^3$ can be represented in 
  the following equivalent form 
   \begin{eqnarray}
  \label{e1.14}
  && \omega_A(B)=B\sqrt{\frac{R}{3M}}
  \end{eqnarray}
  In what follows we use the standard parametrization of
  the nucleus mass $M=m\,A$ (where $m=1.66 \times 10^{-24}$ g is the atomic 
  unit of the nucleon mass) and radius $R=r_0\,A^{1/3}$ (with $r_0=1.2\,10^{-13}$ 
  cm).  On account of this one finds that mass-number dependence of energies (in MeV) of expected hydromagnetic resonant modes 
  is given by  
  \begin{eqnarray}
  \label{e1.15}
  && E_\ell(MHD)=\hbar\omega_\ell(MHD)=\kappa_\ell B\,A^{-1/3},\quad \kappa_\ell=\hbar\left[\frac{r_0}{3 m}\right]^{1/2}\,s_\ell,\\
  \nonumber
  && \kappa_\ell=1.6 \times 10^{-22}\times s_\ell,\quad [2<s_\ell<6,\quad  2<\ell<4]. 
  \end{eqnarray}
   Equation (\ref{e1.15}) exhibits the fact that the nuclear hydromagnetic resonances, if exist, are characterized by one and  the same  
  dependence of energy centroids upon mass number $A$ as empirically established giant resonances
  \begin{eqnarray}
 \label{e1.16}
  && E_{GR}(\ell)=C_\ell\,A^{-1/3}={\rm constant}.
 \end{eqnarray} 
  This suggests, if some of detected giant resonances are of predominantly hydromagnetic nature, 
  then from identification of above theoretical and experimental estimates,  
  $E_\ell(MHD)A^{1/3}=E_{GR}(\ell)A^{1/3}$, one can evaluate the 
  magnitude of nuclear internal magnetic field $B$. 
  For the giant resonant modes lying in the energy interval   
   \begin{eqnarray}
   \label{e3.5}    
  && 30<E_{\rm 
  GR}A^{1/3}<90,\quad (5\leq E_{GR}\leq 15)\,\,\,{\rm [MeV]},\\
 \label{e3.5a}    
  && 4.8 \times 10^{-5}<E_{\rm GR}A^{1/3}<1.4\times 10^{-4}\,\,\,\,{\rm [erg]}
   \end{eqnarray} 
   we obtain  
  \begin{eqnarray}
\label{e3.6}
&& 3.0\times 10^{17} \leq B \leq  9.0\times 10^{17}\,{\rm G}.
\end{eqnarray}
 Remarkably, that according to QCD estimates of radius distribution of magnetic moment in  a nucleon  (whose origin is attributed to 
 persistent Fermi-motion of quarks) falls in the range $0.3<R_N<0.6$ fm. 
 Taking into account that nuclear magneton, $\mu_N=5.05\times 10^{-24}$ erg/G, it easy to see that the intensity of dipole magnetic field 
 $B=2\mu_N/R_N^3$ on the magnetic poles of sphere of above radius 
 (spherical region occupied by quark matter) is ranged in the interval: $5.0\times 10^{16} - 5.0 \times  10^{17}$ G. 
 Similar argument has been used in paper\cite{ALP-96}, devoted to the 
 possibility of ferromagnetic state of superdense matter. The above estimates give an idea about magnetic field intensity in 
 the quark matter which is expected to exist in deep cores of neutrons stars\cite{Xu-09}.
 In this latter context also noteworthy that current investigations on search for the chiral magnetic effects
 lead to the conclusion that magnetic fields of above strength should be generated  in heavy-ion collisions
 at intermediate energies\cite{SIT-09,OB-11}.
 It seems interesting to note that the magnetic field energy stored in spherical 
 volume of nuclear radius, $R=r_0A^{1/3}$,  is proportional to the mass number $A$, namely: $W_B\sim B^2R^3\sim\,A$, as is the case 
 of volume-energy-term in semi-empirical formula for the nuclear binding  energy. 
 This suggests that the volume term of nuclear binding energy 
 may be of magnetic origin, that is, due to the energy of huge magnetic field stored 
 in the nucleus on the stage of explosive nucleosynthesis.  In this connction it is appropriate to note that synthesis of chemical   
 elements in the presence of a super strong magnetic fields of magnetars has recently been studied in 
  [28-31] with remarkable conclusion 
  that the fields of order of $B\sim 10^{17}$ G can substantially affect 
  both the r-process of neutron capture and formation of shell nuclear structure, that 
  is, magic nuclei with enhanced stability. Finally, it may be worth 
  mentioning the well-known in astrophysics argument\cite{SW-61} regarding the effect of 
  strong internal magnetic field on the star shape: 
  the prevailed poloidal magnetic field leads to the oblate deformation of the star 
  shape, whereas the toroidal field leads to prolate deformation\footnote{It may be worth noting that this paper 
  of Sweet [32] is the one in which the conservation of magnetic flux density in the process of the main-sequence star formation from 
  gravitationally contructing gas-dust interstellar medium has been discussed for the first time. 
  As is commonly known today, Ginzburg and Woltjer were the first 
  to suggest that super strong magnetic fields of neutron stars can too be explained as due to the magnetic flux conservation in process 
  of gravitation collapse of massive main-sequence stars.}.
  From this perspective, it is not implausible to expect, therefore, that 
  it is the super strong internal magnetic field plays decisive part in the formation of equilibrium shapes of nuclei heavier 
  than Fe-56.

\section{Summary}

  As a development of Alfv\'en hypothesis about the atomic nucleus capability of responding by magneto-hydrodynamic 
  vibrations, we have set up a collective model providing theoretical basis for computing 
  energies of nuclear hydromagnetic resonances.
   The central to this model, which is appropriate to nuclei heavier than Fe-56, 
  is the intranuclear magnetic field. This  field is considered as being frozen-in the  mid-wight and heavy nuclei  
  on the stage of their formation in the r-process of explosive nucleosynthesis.
  The model predicts that the energy of nuclear hydromagnetic resonances is a linear 
  function of internal magnetic field. The mass-number dependence of energy has one and the 
  same shape as that for typical giant resonances. Based on this and assuming that 
  some of observed giant resonances are predominantly of hydromagnetic nature we found 
  that the intensity of intranuclear magnetic field falls in the realm of magnetic fields of magnetars.

 \section*{Acknowledgments}
 
  This work was supported by the National Natural Science Foundation of China
(Grant Nos. 10935001, 10973002), the National Basic Research Program of
China (Grant No. 2009CB824800), and the John Templeton Foundation.

  \end{document}